# Applicability of the interface spring model for micromechanical analyses with interfacial imperfections to predict the modified exterior Eshelby tensor and effective modulus

Sangryun Lee[1], Youngsoo Kim[1], Jinyeop Lee[2], and Seunghwa Ryu[1,*]

**Affiliations**

[1]Department of Mechanical Engineering and [2]Department of Mathematical Sciences, Korea Advanced Institute of Science and Technology (KAIST), 291 Daehak-ro, Yuseong-gu, Daejeon 34141, Republic of Korea

[*]Corresponding author e-mail: ryush@kaist.ac.kr

**Abstract**

Closed-form solutions for the modified exterior Eshelby tensor, strain concentration tensor, and effective moduli of particle-reinforced composites are presented when the interfacial damage is modeled as a linear spring layer of vanishing thickness; the solutions are validated against finite element analyses. Based on the closed-form solutions, the applicability of the interface spring model is tested by calculating those quantities using finite element analysis (FEA) augmented with a matrix–inhomogeneity non-overlapping condition. The results indicate that the interface spring model reasonably captures the characteristics of the stress distribution and effective moduli of composites, despite its well-known problem of unphysical overlapping between the matrix and inhomogeneity.

## 1. Introduction

The mechanical properties of multicomponent alloys and composite structures can be deduced by considering the strain fields in the inclusions and inhomogeneity. An embedded material having an eigenstrain with an identical elastic stiffness tensor, $\boldsymbol{L}_0$, as the matrix is referred to as an inclusion, while a material with a different elastic stiffness tensor, $\boldsymbol{L}_1$, is referred to as an inhomogeneity. The eigenstrain refers to the stress-free deformation strain of the free-standing inclusion, which can originate from thermal expansion, initial strain, or phase transformation[1-4]. In 1957, J. D. Eshelby proved that the strain field within an ellipsoidal inclusion embedded in an infinite matrix is uniform when the inclusion is subject to a uniform eigenstrain[5]. The constrained strain, $\boldsymbol{\varepsilon}^{\mathrm{c}}$, inside the inclusion embedded in the matrix is related to the eigenstrain, $\boldsymbol{\varepsilon}^{*}$, via the interior Eshelby tensor, $\boldsymbol{S}^{\mathrm{Int,P}}$, as follows: $\boldsymbol{\varepsilon}^{\mathrm{c}} = \boldsymbol{S}^{\mathrm{Int,P}}: \boldsymbol{\varepsilon}^{*}$, where the superscripts "Int" and "P" indicate the interior of the inclusion and a perfect matrix–inclusion interface, respectively; and ":" refers to the double contraction operation[5, 6].

The strain field inside an ellipsoidal inhomogeneity subject to an external load can be obtained by transforming the problem into the corresponding equivalent inclusion problem[7, 8]. Mean field micromechanics models such as the Mori–Tanaka method utilize the Eshelby tensor, $\boldsymbol{S}^{\mathrm{Int,P}}$, to predict the effective properties of a composite by relating the average macroscopic external strain, $\boldsymbol{\varepsilon}^{0}$, to the average internal strain field inside the inhomogeneity, $\boldsymbol{\varepsilon}^{1}$, as $\boldsymbol{\varepsilon}^{1} = \boldsymbol{A}^{\mathrm{Int,P}}: \boldsymbol{\varepsilon}^{0}$, where $\boldsymbol{A}^{\mathrm{Int,P}}$ is the interior strain concentration tensor and is given in terms of the interior Eshelby tensor, $\boldsymbol{S}^{\mathrm{Int,P}}$, and elastic stiffness tensors, $\boldsymbol{L}_0$ and $\boldsymbol{L}_1$, of the matrix and the inhomogeneity, respectively[8]. Hence, given the importance of the Eshelby tensor, a primary focus of recent research has been the derivation of the simplified form of $\boldsymbol{S}^{\mathrm{Int,P}}$ for various material properties and geometries of the inclusion[9-11].

The strain field of the matrix around the inclusion has also been studied because the strain field outside of the inclusion plays a critical role in determining the mechanical properties beyond the elastic response regime. It is also important for considering composites with high reinforcement volume fractions, in which the interaction among the reinforcements becomes important. The non-uniform strain field at a position $\boldsymbol{x}$ in the matrix can be expressed with a similar form as $\boldsymbol{\varepsilon}^{\mathrm{c}}(\boldsymbol{x}) = \boldsymbol{S}^{\mathrm{Ext,P}}(\boldsymbol{x})\!:\boldsymbol{\varepsilon}^{*}$, where the superscript "Ext" refers to the exterior, and $\boldsymbol{S}^{\mathrm{Ext,P}}(\boldsymbol{x})$ is the exterior Eshelby tensor[12, 13]. The exterior Eshelby tensor and the exterior strain concentration tensor, $\boldsymbol{A}^{\mathrm{Ext,P}}(\boldsymbol{x})$, have been used to consider the shear localization of metallic glass, the matrix failure point for reinforced composites, and the interaction among the reinforcements[14-16].

Further progress on the inclusion or inhomogeneity problem has been achieved by considering the interfacial damage present in realistic composites where debonding or slip between the matrix and inclusion can occur. As a simplified model of interfacial imperfection, J. Qu introduced a linear-spring layer of vanishing thickness at the interface to represent the displacement jump between the inclusion and the matrix[17, 18]. Zero and infinite interfacial spring compliance correspond to perfect and completely damaged (i.e., no load transfer between the matrix and the inclusion) interfaces, respectively. Owing to its mathematical simplicity and easier physical interpretation (compared to the interface stress model[19, 20] and the interphase model[21, 22]), the interface spring model has been widely adopted to describe composites with an imperfect interface[17, 18, 23-26]. In earlier studies, the modified interior Eshelby tensor, $\boldsymbol{S}^{\mathrm{Int,D}}$, and the modified interior strain concentration tensor, $\boldsymbol{A}^{\mathrm{Int,D}}$, were obtained and applied to predict the effective mechanical properties of composites with interfacial damage[24, 25, 27]. Here, the superscript "D" refers to a damaged interface. However, because of mathematical errors in these earlier studies, such as violation of the Fubini-Tonelli theorem, the predicted tensors have a nonphysical singular value at some specific interface spring compliance values[7, 28]. A recent study by Y. Othmani et al. discussed these errors in detail and presented correct solutions for $\boldsymbol{S}^{\mathrm{Int,D}}$ and $\boldsymbol{A}^{\mathrm{Int,D}}$ for a spherical inclusion[7]. Our previous paper reported a simple tensor algebraic form of $\boldsymbol{S}^{\mathrm{Int,D}}$ for a spherical inclusion embedded in an arbitrarily anisotropic matrix[29]. Similar considerations can be extended to the thermal conductivity of composites within an anisotropic matrix with interfacial imperfections (e.g., Kapitza resistance) based on the mathematical analogy[30].

This study further explores this line of investigation for two aspects of the modified Eshelby tensor and the effective stiffness of composites in the presence of interfacial damage. First, to the best of our knowledge, there has been no consideration of the modified exterior Eshelby tensor, $\boldsymbol{S}^{\mathrm{Ext,D}}$, and the modified exterior strain concentration tensor, $\boldsymbol{A}^{\mathrm{Ext,D}}$. This study derives closed-form solutions for $\boldsymbol{S}^{\mathrm{Ext,D}}$ and $\boldsymbol{A}^{\mathrm{Ext,D}}$ which are applicable to an isotropic matrix; the solutions are validated against the predictions of finite element analysis (FEA) over a wide range of interface spring compliance values (i.e., degrees of imperfection). Second, we systematically examine a weakness of the interface spring model, i.e., the unphysical overlap between the matrix and the inclusion, which arises from the simplified description based on a linear interfacial spring[19]. The modified exterior stress tensor, $\boldsymbol{B}^{\mathrm{Ext,D}}$, and the effective moduli of composites obtained with three methods (the interface spring theory, FEA with an interface spring, and FEA with an interface spring augmented with a non-overlapping matrix–inhomogeneity contact constraint) are compared. The results show that the effect of unphysical overlapping on the prediction of effective stiffness is relatively small when the volume fraction is low, but becomes noticeable when the fraction increases. This study deepens the understanding of the micromechanics of composites in the presence of interfacial damage and clarifies the applicability of the interfacial spring model.

## 2. Theory

### 2.1 Single inclusion problem in the absence of interfacial damage

In elastostatics, Green's function ($G_{kp}(\boldsymbol{x}-\boldsymbol{y})$, expressing displacement in the $k$-direction under a unit body force in the $p$-direction) can be obtained from the following governing equation:

$$L_{ijkl}G_{kp,lj}(\boldsymbol{x}-\boldsymbol{y})+\delta_{ip}\delta(\boldsymbol{x}-\boldsymbol{y})=0 \tag{1}$$

where $L_{ijkl}$ is a fourth-order stiffness tensor, and $\blacksquare_{,i}$ implies $\partial\blacksquare/\partial x_i$. Hereafter, the repeated indices are dummy indices which imply the summation of all values, 1 to 3. Green's function for an isotropic material can be expressed as follows:

$$G_{ij}(\boldsymbol{x}-\boldsymbol{y})=\frac{1}{16\pi\mu(1-\nu)|\boldsymbol{x}-\boldsymbol{y}|}\left[(3-4\nu)\delta_{ij}+\frac{(x_i-y_i)(x_j-y_j)}{|\boldsymbol{x}-\boldsymbol{y}|^2}\right] \tag{2}$$

where $\mu$ and $\nu$ are the shear modulus and Poisson's ratio of the material, respectively. Green's function for an anisotropic material can be expressed in either a closed form or using series expansion[9, 31].

The single inclusion problem proposed by J.D. Eshelby can be depicted as shown in Figure 1. The inclusion is deformed by the eigenstrain, $\boldsymbol{\varepsilon}^*$, in the absence of any constraint, and is then inserted into the hole while subjected to a traction, $\boldsymbol{T}$, to maintain the original shape. Upon the release of the traction, $\boldsymbol{T}$, the inclusion will expand owing to the constrained strain, $\boldsymbol{\varepsilon}^C$, which differs from $\boldsymbol{\varepsilon}^*$ due to the interaction with the matrix. The constrained strain field can be expressed as follows:

$$\varepsilon_{ij}^{c}(\boldsymbol{x}) = \left[\frac{1}{2}\int_{\Omega} L_{pqrs}\left\{\frac{\partial^2 G_{ip}(\boldsymbol{x}-\boldsymbol{y})}{\partial x_j \partial y_q} + \frac{\partial^2 G_{jp}(\boldsymbol{x}-\boldsymbol{y})}{\partial x_i \partial y_q}\right\} d\boldsymbol{y}\right] \varepsilon_{rs}^{*}.$$

$$\varepsilon_{ij}^{c}(\boldsymbol{x}) = S_{ijrs}^{\mathrm{P}}(\boldsymbol{x})\varepsilon_{rs}^{*} = \begin{cases} S_{ijrs}^{\mathrm{Int,P}}\varepsilon_{rs}^{*} & \boldsymbol{x} \in \Omega \\ S_{ijrs}^{\mathrm{Ext,P}}(\boldsymbol{x})\varepsilon_{rs}^{*} & \boldsymbol{x} \in D \setminus \Omega \end{cases} \tag{3}$$

where $\Omega$ is the inclusion volume, and $D$ is the total volume including the matrix and inclusion[5, 32].

For a spherical inclusion in an isotropic matrix with a perfect interface, the closed-form solutions for the interior Eshelby tensor and exterior Eshelby tensor can be derived as Eqs. (4) and (5), respectively[13].

$$S_{ijrs}^{\mathrm{Int,P}} = \frac{(5\nu-1)}{15(1-\nu)}\delta_{ij}\delta_{rs} + \frac{(4-5\nu)}{15(1-\nu)}\left(\delta_{ir}\delta_{js} + \delta_{jr}\delta_{is}\right) \tag{4}$$

$$\begin{aligned} S_{ijrs}^{\mathrm{Ext,P}}(\boldsymbol{x}) = \frac{\rho^3}{30(1-\nu)} \big[ & (3\rho^2 + 10\nu - 5)\delta_{ij}\delta_{rs} \\ & + (3\rho^2 + 10\nu - 5)\left(\delta_{ir}\delta_{js} + \delta_{is}\delta_{jr}\right) + 15(1-\rho^2)\delta_{ij}\bar{x}_r\bar{x}_s \\ & + 15(1 - 2\nu - \rho^2)\delta_{rs}\bar{x}_i\bar{x}_j \\ & + 15(\nu - \rho^2)\left(\delta_{ir}\bar{x}_j\bar{x}_s + \delta_{jr}\bar{x}_i\bar{x}_s + \delta_{is}\bar{x}_j\bar{x}_r + \delta_{js}\bar{x}_i\bar{x}_r\right) \\ & + 15(7\rho^2 - 5)\bar{x}_i\bar{x}_j\bar{x}_r\bar{x}_s \big] \end{aligned} \tag{5}$$

where $\rho = |\boldsymbol{x}|/R$ and $\bar{x}_i = x_i/R$ are the distance and $i^{\mathrm{th}}$ coordinate from the center of the inclusion, respectively, normalized with respect to the inclusion radius, $R$. The interior and exterior Eshelby tensors for an anisotropic material or different inclusion shapes with a perfect interface have been derived in previous studies[12, 33, 34].

## 2.2 Single inclusion problem in the presence of interfacial damage

The interface spring model considers the interfacial imperfection as a linear spring with vanishing thickness (see Figure 2)[35]. Zero and infinite interfacial spring compliance correspond to perfect and completely damaged (i.e., no load transfer between the matrix and the inclusion) interfaces, respectively. The traction equilibrium equation and the constitutive equation at the interface are expressed as follows:

$$\begin{aligned} \Delta t_i = \Delta\sigma_{ij}n_j = \left[\sigma_{ij}(\partial\Omega^+) - \sigma_{ij}(\partial\Omega^-)\right]n_j = 0 \\ \Delta u_i = \left[u_i(\partial\Omega^+) - u_i(\partial\Omega^-)\right] = \eta_{ij}\sigma_{jk}n_k \end{aligned} \tag{6}$$

where $(\partial\Omega^+)$ and $(\partial\Omega^-)$ denote the interface at the matrix and inclusion sides, respectively. Note that $\Delta u_i$ represents the displacement jump across the interface, and $\boldsymbol{n}$ is the unit outward normal vector at the inclusion surface. The term $\eta_{ij}$ is the second-order spring compliance tensor, which can be expressed in terms of the tangential spring compliance, $\alpha$, and normal spring compliance, $\beta$, as follows:

$$\eta_{ij} = \alpha\delta_{ij} + (\beta - \alpha)n_i n_j \tag{7}$$

By adopting these interfacial conditions, the constrained strain in the modified Eshelby inclusion problem can be written as follows:

$$\begin{aligned} \varepsilon_{ij}^{c}(\boldsymbol{x}) = {} & S_{ijrs}^{\mathrm{P}}(\boldsymbol{x})\varepsilon_{rs}^{*} \\ & + \frac{1}{2}L_{klmn}L_{pqrs}\int_{\partial\Omega}\left[\eta_{kp}\left\{\frac{\partial^2 G_{im}(\boldsymbol{x}-\boldsymbol{y})}{\partial x_j \partial y_n}\right.\right. \\ & \left.\left. + \frac{\partial^2 G_{jm}(\boldsymbol{x}-\boldsymbol{y})}{\partial x_i \partial y_n}\right\} n_q(\boldsymbol{y})n_l(\boldsymbol{y})\left(\varepsilon_{rs}^{c}(\boldsymbol{y}) - \varepsilon_{rs}^{*}\right)\right] d\boldsymbol{y}. \end{aligned} \tag{8}$$

Because the strain field within the inclusion is generally non-uniform (expressed as a quadratic function of $\bar{x}_i$ for the spherical inclusion)[26, 36] in the presence of interfacial damage, it is difficult to obtain the relationship between $\varepsilon_{ij}^{c}$ and $\varepsilon_{rs}^{*}$ from the implicit integral equation involving $\varepsilon_{rs}^{c}$. However, Z. Zhong and S. A. Meguid found that the strain field within the inclusion is uniform if two conditions are met[26]: first, the two compliances are the same $(\alpha = \beta \equiv \gamma)$ so that the constitutive equation becomes $\Delta u_i = \gamma t_i$, which is a similar form to Hooke's law; and second, the inclusion shape is perfectly spherical. In a special case of this type, $\varepsilon_{rs}^{c}$ can be taken out of the integral on the right-hand side as follows:

$$\begin{aligned}\varepsilon_{ij}^{c}(\boldsymbol{x}) &= S_{ijrs}^{\mathrm{P}}(\boldsymbol{x})\varepsilon_{rs}^{*}\\ &\quad + \frac{1}{2}\gamma L_{klmn}L_{kqrs}\int_{\partial\Omega}\left[\left\{\frac{\partial^2 G_{im}(\boldsymbol{x}-\boldsymbol{y})}{\partial x_j \partial y_n}\right.\right.\\ &\quad \left.\left. + \frac{\partial^2 G_{jm}(\boldsymbol{x}-\boldsymbol{y})}{\partial x_i \partial y_n}\right\} n_q(\boldsymbol{y}) n_l(\boldsymbol{y})\right] d\boldsymbol{y}\left(\overline{\varepsilon_{rs}^{c}} - \varepsilon_{rs}^{*}\right).\end{aligned} \tag{9}$$

The term $\overline{\varepsilon_{rs}^{c}}$ is the strain within the inclusion, and it can be expressed as $\overline{\varepsilon_{rs}^{c}} = S_{rskl}^{\mathrm{Int,D}}\varepsilon_{kl}^{*}$, where $S_{rskl}^{\mathrm{Int,D}}$ is the modified interior Eshelby tensor. If $\Gamma_{ijrs}(\boldsymbol{x})$ is defined such that the following is the case:

$$\Gamma_{ijrs}(\boldsymbol{x}) \equiv -\frac{1}{2}\gamma L_{klmn}L_{kqrs}\int_{\partial\Omega}\left[\left\{\frac{\partial^2 G_{im}(\boldsymbol{x}-\boldsymbol{y})}{\partial x_j \partial y_n} + \frac{\partial^2 G_{jm}(\boldsymbol{x}-\boldsymbol{y})}{\partial x_i \partial y_n}\right\} n_q(\boldsymbol{y}) n_l(\boldsymbol{y})\right] d\boldsymbol{y}. \tag{10}$$

then the constrained strain field for point $\boldsymbol{x} \in \mathbb{R}^3$ can be rewritten as follows:

$$\varepsilon_{ij}^{c}(\boldsymbol{x}) = S_{ijrs}^{\mathrm{P}}(\boldsymbol{x})\varepsilon_{rs}^{*} - \Gamma_{ijrs}(\boldsymbol{x})\left(\overline{\varepsilon_{rs}^{c}} - \varepsilon_{rs}^{*}\right) \tag{11}$$

For $\boldsymbol{x} \in \Omega$, $\varepsilon_{ij}^{c}(\boldsymbol{x}) = \overline{\varepsilon_{ij}^{c}}$ , $\Gamma_{ijrs}(\boldsymbol{x}) = \bar{\Gamma}_{ijrs}$, and $S_{ijrs}^{\mathrm{P}}(\boldsymbol{x}) = S_{ijrs}^{\mathrm{Int,P}}$ . Therefore, the strain field within the inclusion and the modified interior Eshelby tensor can be obtained as follows:

$$\overline{\boldsymbol{\varepsilon}^{c}} = (\boldsymbol{I} + \overline{\boldsymbol{\Gamma}})^{-1} : (\boldsymbol{S} + \overline{\boldsymbol{\Gamma}}) : \boldsymbol{\varepsilon}^{*} = \boldsymbol{S}^{\mathrm{Int,D}} : \boldsymbol{\varepsilon}^{*} \tag{12}$$

$$\boldsymbol{S}^{\mathrm{Int,D}} = (\boldsymbol{I} + \overline{\boldsymbol{\Gamma}})^{-1} : (\boldsymbol{S} + \overline{\boldsymbol{\Gamma}}) \tag{13}$$

where $I_{ijkl} = \frac{1}{2}\left(\delta_{ik}\delta_{jl} + \delta_{il}\delta_{jk}\right)$ is the fourth-order symmetric identity tensor. $\overline{\boldsymbol{\Gamma}}$ for an arbitrary anisotropic matrix having a spherical inclusion can be written as Eq. (14)[29].

$$\bar{\Gamma}_{ijrs} = \frac{\gamma}{R}\left(I_{ijpq} - S_{ijpq}^{\mathrm{Int,P}}\right)L_{pqrs} \tag{14}$$

For an isotropic matrix with a spherical inclusion, this can be simplified as follows:

$$\bar{\Gamma}_{ijkl} = \frac{\mu\gamma}{15R(1-\nu)}\left[2(1+5\nu)\delta_{ij}\delta_{kl} + (7-5\nu)\left(\delta_{ik}\delta_{jl} + \delta_{il}\delta_{jk}\right)\right] \tag{15}$$

For $\boldsymbol{x} \in D \setminus \Omega$, the strain field within the matrix is expressed as follows:

$$\varepsilon_{ij}^{c}(\boldsymbol{x}) = S_{ijkl}^{\mathrm{P}}(\boldsymbol{x})\varepsilon_{kl}^{*} - \Gamma_{ijrs}(\boldsymbol{x})\left(S_{rskl}^{\mathrm{Int,D}} - I_{rskl}\right)\varepsilon_{kl}^{*} = S_{ijkl}^{\mathrm{Ext,D}}(\boldsymbol{x})\varepsilon_{kl}^{*} \tag{16}$$

where

$$S_{ijkl}^{\mathrm{Ext,D}}(\boldsymbol{x}) \equiv S_{ijkl}^{\mathrm{Ext,P}}(\boldsymbol{x}) - \Gamma_{ijrs}(\boldsymbol{x})\left(S_{rskl}^{\mathrm{Int,D}} - I_{rskl}\right). \tag{17}$$

To predict the constrained strain field, it is essential to calculate $\Gamma_{ijrs}(\boldsymbol{x})$. In this study, the closed-form solution of $\Gamma_{ijrs}(\boldsymbol{x})$ is derived for an isotropic matrix with a spherical inclusion. $G_{ij}(\boldsymbol{x}-\boldsymbol{y})$ in Eq. (2) can be expressed in alternative form as follows:

$$G_{ij}(\boldsymbol{x}-\boldsymbol{y}) = \frac{\delta_{ij}}{4\pi\mu|\boldsymbol{x}-\boldsymbol{y}|} - \frac{1}{16\pi\mu(1-\nu)}\frac{\partial^2}{\partial x_i \partial x_j}|\boldsymbol{x}-\boldsymbol{y}|. \tag{18}$$

Given that $\frac{\partial^2 G_{im}(\boldsymbol{x}-\boldsymbol{y})}{\partial x_j \partial y_n} = -\frac{\partial^2 G_{im}(\boldsymbol{x}-\boldsymbol{y})}{\partial x_j \partial x_n}$, Eq. (10) can be written as follows:

$$\Gamma_{ijrs}(\boldsymbol{x}) = \frac{1}{2}\gamma L_{klmn}L_{kqrs}\int_{\partial\Omega}\left[\left\{G_{im,jn}(\boldsymbol{x}-\boldsymbol{y}) + G_{jm,in}(\boldsymbol{x}-\boldsymbol{y})\right\}n_q(\boldsymbol{y})n_l(\boldsymbol{y})\right]d\boldsymbol{y}. \tag{19}$$

After applying the divergence theorem by noting that $n_q(\boldsymbol{y}) = y_q/R$, $\Gamma_{ijrs}(\boldsymbol{x})$ can be written as follows:

$$\varGamma_{ijrs}(\boldsymbol{x}) = -\frac{1}{2}\frac{\gamma}{R}L_{klmn}L_{kqrs}\left[\frac{\partial^2}{\partial x_n \partial x_l}\left(\frac{\partial}{\partial x_j}\int_\Omega G_{im}(\boldsymbol{x}-\boldsymbol{y})y_q d\boldsymbol{y} + \frac{\partial}{\partial x_i}\int_\Omega G_{jm}(\boldsymbol{x}-\boldsymbol{y})y_q d\boldsymbol{y}\right) + \delta_{ql}\frac{\partial}{\partial x_n}\left(\frac{\partial}{\partial x_j}\int_\Omega G_{im}(\boldsymbol{x}-\boldsymbol{y})d\boldsymbol{y} + \frac{\partial}{\partial x_i}\int_\Omega G_{jm}(\boldsymbol{x}-\boldsymbol{y})d\boldsymbol{y}\right)\right] \tag{20}$$

Plugging Eq. (18) into Eq. (20) yields the following expression:

$$\varGamma_{ijrs}(\boldsymbol{x}) = -\frac{1}{2}\frac{\gamma}{R}L_{klmn}L_{kqrs}\left[\frac{1}{4\pi\mu}\left(\delta_{mi}J_{q,jln}(\boldsymbol{x}) + \delta_{mj}J_{q,iln}(\boldsymbol{x}) + \delta_{ql}\delta_{im}\phi_{,jn}(\boldsymbol{x}) + \delta_{ql}\delta_{jm}\phi_{,in}(\boldsymbol{x})\right) - \frac{1}{8\pi\mu(1-\nu)}\left(I_{q,ijlmn}(\boldsymbol{x}) + \delta_{ql}\psi_{,ijmn}(\boldsymbol{x})\right)\right] \tag{21}$$

where $\phi(\boldsymbol{x})$, $\psi(\boldsymbol{x})$, $J_i(\boldsymbol{x})$, and $I_i(\boldsymbol{x})$ can be obtained from potential theory and written as follows[9, 37, 38]:

$$\phi(\boldsymbol{x}) = \int_\Omega \frac{1}{|\boldsymbol{x}-\boldsymbol{y}|}d\boldsymbol{y} = \frac{4\pi R^3}{3|\boldsymbol{x}|},$$

$$\psi(\boldsymbol{x}) = \int_\Omega |\boldsymbol{x}-\boldsymbol{y}|d\boldsymbol{y} = \frac{4\pi R^3}{3}\left(|\boldsymbol{x}| + \frac{R^2}{5|\boldsymbol{x}|}\right),$$

$$J_i(\boldsymbol{x}) = \int_\Omega \frac{y_i}{|\boldsymbol{x}-\boldsymbol{y}|}d\boldsymbol{y} = \frac{4\pi R^5}{15|\boldsymbol{x}|^3}x_i, \tag{22}$$

and

$$I_i(\boldsymbol{x}) = \int_\Omega y_i|\boldsymbol{x}-\boldsymbol{y}|d\boldsymbol{y} = \frac{4\pi R^7}{105|\boldsymbol{x}|^3}x_i - \frac{4\pi R^5}{15|\boldsymbol{x}|}x_i.$$

Combining Eqs. (21) and (22) yields a simplified expression for $\varGamma_{ijrs}(\boldsymbol{x})$:

$$
\begin{aligned}
\Gamma_{ijrs}(\boldsymbol{x}) = \frac{\rho^3\mu\gamma}{R}\Big[&\frac{1}{3}\big(\delta_{ij}\delta_{rs} - 2\delta_{ir}\delta_{js} - 2\delta_{is}\delta_{jr} - 3\delta_{rs}\bar{x}_i\bar{x}_j + 3\delta_{js}\bar{x}_i\bar{x}_r + 3\delta_{is}\bar{x}_j\bar{x}_r \\
&+ 3\delta_{jr}\bar{x}_i\bar{x}_s + 3\delta_{ir}\bar{x}_j\bar{x}_s\big) \\
&+ \frac{1}{15(1-\nu)}\big(15\delta_{ij}\delta_{rs} + 5\delta_{ir}\delta_{js} + 5\delta_{is}\delta_{jr} - 45\delta_{rs}\bar{x}_i\bar{x}_j - 15\delta_{js}\bar{x}_i\bar{x}_r \\
&- 15\delta_{is}\bar{x}_j\bar{x}_r - 15\delta_{jr}\bar{x}_i\bar{x}_s - 15\delta_{ir}\bar{x}_j\bar{x}_s - 15\delta_{ij}\bar{x}_r\bar{x}_s - 3\rho^2\delta_{ij}\delta_{rs} \\
&- 3\rho^2\delta_{ir}\delta_{js} - 3\rho^2\delta_{is}\delta_{jr} - 105\rho^2\bar{x}_i\bar{x}_j\bar{x}_r\bar{x}_s + 75\bar{x}_i\bar{x}_j\bar{x}_r\bar{x}_s \\
&+ 15\rho^2\delta_{rs}\bar{x}_i\bar{x}_j + 15\rho^2\delta_{js}\bar{x}_i\bar{x}_r + 15\rho^2\delta_{is}\bar{x}_j\bar{x}_r + 15\rho^2\delta_{jr}\bar{x}_i\bar{x}_s \\
&+ 15\rho^2\delta_{ir}\bar{x}_j\bar{x}_s + 15\rho^2\delta_{ij}\bar{x}_r\bar{x}_s\big) + \frac{1}{(1-2\nu)}\big(3\delta_{rs}\bar{x}_i\bar{x}_j - \delta_{ij}\delta_{rs}\big)\Big].
\end{aligned} \tag{23}
$$

Although Eq. (23) appears complicated, from a computational standpoint, it is much simpler than using numerical approaches such as FEA.

As shown in Eqs. (13) and (17), it is necessary to conduct inverse operations and double inner products of the fourth-order tensor to predict the interior and exterior modified Eshelby tensors. In this paper, Mandel notation has been adopted to transform the fourth-order tensor into a $6 \times 6$ matrix[38]. As discussed in our previous study[29], the fourth-order tensor, $\boldsymbol{A}, \boldsymbol{B}$, with minor symmetry ($A_{ijkl} = A_{jikl} = A_{ijlk}$) can be transformed into a $6 \times 6$ matrix, $\langle \boldsymbol{A} \rangle, \langle \boldsymbol{B} \rangle$, as follows:

$$
\langle \boldsymbol{A} \rangle = \begin{bmatrix}
A_{1111} & A_{1122} & A_{1133} & \sqrt{2}A_{1123} & \sqrt{2}A_{1131} & \sqrt{2}A_{1112} \\
A_{2211} & A_{2222} & A_{2233} & \sqrt{2}A_{2223} & \sqrt{2}A_{2231} & \sqrt{2}A_{2212} \\
A_{3311} & A_{3322} & A_{3333} & \sqrt{2}A_{3323} & \sqrt{2}A_{3331} & \sqrt{2}A_{3312} \\
\sqrt{2}A_{2311} & \sqrt{2}A_{2322} & \sqrt{2}A_{2333} & 2A_{2323} & 2A_{2331} & 2A_{2312} \\
\sqrt{2}A_{3111} & \sqrt{2}A_{3122} & \sqrt{2}A_{3133} & 2A_{3123} & 2A_{3131} & 2A_{3112} \\
\sqrt{2}A_{1211} & \sqrt{2}A_{1222} & \sqrt{2}A_{1233} & 2A_{1223} & 2A_{1231} & 2A_{1212}
\end{bmatrix}. \tag{24}
$$

The inverse and double inner product can then be conducted using matrix calculation with the following relation:

$$\langle \boldsymbol{A} : \boldsymbol{B} \rangle = \langle \boldsymbol{A} \rangle \langle \boldsymbol{B} \rangle, \qquad \langle \boldsymbol{A}^{-1} \rangle = \langle \boldsymbol{A} \rangle^{-1} \tag{25}$$

For example, the modified interior Eshelby tensor can be predicted using Mandel notation, and is expressed as follows[29]:

$$\langle \boldsymbol{S}^{\mathrm{Int,D}} \rangle = (\langle \boldsymbol{I} \rangle + \langle \bar{\boldsymbol{\Gamma}} \rangle)^{-1} (\langle \boldsymbol{S}^{\mathrm{Int,P}} \rangle + \langle \bar{\boldsymbol{\Gamma}} \rangle). \tag{26}$$

## 2.3. Single inhomogeneity problem in the presence of interfacial damage subject to external loading

The strain field within an inhomogeneity embedded in an infinite matrix under an applied load can be predicted using the equivalent inclusion method. As demonstrated in the literature, in the presence of an interfacial spring, the problem can be decomposed into a linear combination of three independent problems, as depicted in Figure 3[26]. Sub-problem I considers the homogeneous material under applied strain, in which the strain field across the entire domain is uniform and equivalent to the applied strain as follows:

$$\boldsymbol{\varepsilon}^{\mathrm{I}}(\boldsymbol{x}) = \boldsymbol{\varepsilon}^{0}. \tag{27}$$

Sub-problem II considers the perturbation arising from the existence of the inhomogeneity. The perturbation can be expressed as an Eshelby tensor with an equivalent eigenstrain[7, 39]. The equivalent eigenstrain matches the stress field within the inhomogeneity and corresponding inclusion, and thus can be expressed as follows:

$$\boldsymbol{\varepsilon}^{\mathrm{II}}(\boldsymbol{x}) = \boldsymbol{S}^{\mathrm{P}}(\boldsymbol{x}) : \boldsymbol{\varepsilon}^{\mathrm{Eq}} \tag{28}$$

$$\boldsymbol{L}_1 : \boldsymbol{\varepsilon}^{\mathrm{c}} = \boldsymbol{L}_0 : (\boldsymbol{\varepsilon}^{\mathrm{c}} - \boldsymbol{\varepsilon}^{\mathrm{Eq}}) \tag{29}$$

where $\boldsymbol{\varepsilon}^{\mathrm{Eq}}$ is the equivalent eigenstrain, which can be obtained with the following:

$$\boldsymbol{\varepsilon}^{\mathrm{Eq}} = -[(\boldsymbol{L}_1 - \boldsymbol{L}_0):\boldsymbol{S}^{\mathrm{Int,P}} + \boldsymbol{L}_0]^{-1}:(\boldsymbol{L}_1 - \boldsymbol{L}_0):\boldsymbol{\varepsilon}^0. \tag{30}$$

Sub-problem III considers the displacement jump at the interface. The strain field due to the displacement jump can be written as Eq. (31)[7].

$$\varepsilon^{\mathrm{III}}(\boldsymbol{x}) = -\boldsymbol{\Gamma}(\boldsymbol{x}):\left(\overline{\boldsymbol{\varepsilon}^{\mathbf{1}}} - \boldsymbol{\varepsilon}^{\mathrm{Eq}}\right) \tag{31}$$

where $\overline{\boldsymbol{\varepsilon}^{\mathbf{1}}}$ is the strain field within the inhomogeneity. For the case of perfect bonding, there are only two sub-problems because the solution to sub-problem III is zero. After superposing the three solutions, the strain field can be expressed as Eq. (32).

$$\begin{aligned}\boldsymbol{\varepsilon}(\boldsymbol{x}) &= \boldsymbol{\varepsilon}^{\mathrm{I}}(\boldsymbol{x}) + \boldsymbol{\varepsilon}^{\mathrm{II}}(\boldsymbol{x}) + \boldsymbol{\varepsilon}^{\mathrm{III}}(\boldsymbol{x}) \\ &= \boldsymbol{\varepsilon}^0 + \boldsymbol{S}^{\mathrm{P}}(\boldsymbol{x}):\boldsymbol{\varepsilon}^{\mathrm{Eq}} - \boldsymbol{\Gamma}(\boldsymbol{x}):\left(\overline{\boldsymbol{\varepsilon}^{\mathbf{1}}} - \boldsymbol{\varepsilon}^{\mathrm{Eq}}\right)\end{aligned} \tag{32}$$

Plugging Eqs. (27) – (31) into Eq. (32) yields the following expression for the modified exterior strain concentration tensor:

$$\begin{aligned}\boldsymbol{\varepsilon}(\boldsymbol{x}) &= \left(\boldsymbol{I} + \boldsymbol{S}^{\mathrm{Ext,P}}(\boldsymbol{x}):\boldsymbol{L}_{\mathbf{0}}^{-\mathbf{1}}:(\boldsymbol{L}_{\mathbf{0}} - \boldsymbol{L}_{\mathbf{1}}):\boldsymbol{A}^{\mathrm{Int,D}} - \boldsymbol{\Gamma}(\boldsymbol{x}):\boldsymbol{L}_{\mathbf{0}}^{-\mathbf{1}}:\boldsymbol{L}_{\mathbf{1}}:\boldsymbol{A}^{\mathrm{Int,D}}\right):\boldsymbol{\varepsilon}^0 \\ &= \boldsymbol{A}^{\mathrm{Ext,D}}(\boldsymbol{x}):\boldsymbol{\varepsilon}^0\end{aligned} \tag{33}$$

When $\boldsymbol{x} \in \Omega$, then $\boldsymbol{A}^{\mathrm{Ext,D}}(\boldsymbol{x}) = \boldsymbol{A}^{\mathrm{Int,D}}$, $\boldsymbol{S}^{\mathrm{Ext,P}}(\boldsymbol{x}) = \boldsymbol{S}^{\mathrm{Int,P}}$, and $\boldsymbol{\Gamma}(\boldsymbol{x}) = \overline{\boldsymbol{\Gamma}}$. Thus, a closed form of $\boldsymbol{A}^{\mathrm{Int,D}}$ can be derived such that

$$\boldsymbol{A}^{\mathrm{Int,D}} = \left(\boldsymbol{I} + \boldsymbol{S}^{\mathrm{Int,P}}:\boldsymbol{L}_{\mathbf{0}}^{-\mathbf{1}}:(\boldsymbol{L}_{\mathbf{1}} - \boldsymbol{L}_{\mathbf{0}}) + \overline{\boldsymbol{\Gamma}}:\boldsymbol{L}_{\mathbf{0}}^{-\mathbf{1}}:\boldsymbol{L}_{\mathbf{1}}\right)^{-1}, \tag{34}$$

which coincides with the result reported in Y. Othmani et al.[7] As shown in Eq. (33), our model can predict the strain field of a matrix with an arbitrary anisotropic spherical inhomogeneity by using an anisotropic $\boldsymbol{L}_1$, whereas the model proposed by Z. Zhong and S. A. Meguid can only be applied to isotropic inhomogeneity problems[26]. The calculations of Eqs. (33) and (34) are also carried out using Mandel notation.

### 2.4. Effective modulus prediction in the presence of interfacial damage subject to external

**loading**

The effective modulus in the presence of interfacial damage can be obtained with the Mori–Tanaka method by correctly accounting for the modified internal Eshelby tensor and modified strain concentration tensor. Existing studies on the prediction of the effective modulus suffer either from an incorrect evaluation of the modified Eshelby tensor due to violation of Fubini's theorem, or from considering a linear superposition of sub-problems that does not match the boundary conditions of the original problem[25, 28, 36, 40-42]. By correctly taking into account the interface spring model, the effective modulus $\left(\boldsymbol{L}^{\mathrm{Eff}}\right)$ of the composite is derived as follows:

$$\boldsymbol{L}^{\mathrm{Eff}} = (c_0\boldsymbol{L}_0 + c_1\boldsymbol{L}_1 : \boldsymbol{A}^{\mathrm{Int,D}}) : \left(c_0\boldsymbol{I} + c_1\boldsymbol{A}^{\mathrm{Int,D}} + c_1\frac{\gamma}{R}\boldsymbol{L}_1 : \boldsymbol{A}^{\mathrm{Int,D}}\right)^{-1} \quad (35)$$

where $c_0$ and $c_1$ are the volume fractions of the matrix and inhomogeneity, respectively, and $c_0 + c_1 = 1$. To the best of our knowledge, this is the first study which uses a correctly calculated modified interior Eshelby tensor and modified strain concentration tensor to predict the effective modulus based on the Mori–Tanaka method.

## 3. Numerical Tests

### 3.1 Finite element analysis

A series of FEA were conducted using COMSOL to validate the modified exterior Eshelby tensor and strain concentration tensor[43]. Because the single inclusion and inhomogeneity problem considers an infinite matrix, a sufficiently large matrix having an edge ten times longer than the diameter (2mm) of the particle is used[29]. The material properties of the matrix and inhomogeneity are $E_0 = 200$ GPa, $\nu_0 = 0.25$ and $E_1 = 400$ GPa, $\nu_1 = 0.25$ respectively. A mesh was constructed for the calculations using approximately 400,000 and 8,000 quadratic tetrahedron elements for the matrix and particle, respectively. The modified Eshelby tensor is obtained from the constrained strain after imposing a unit eigenstrain, and the modified strain concentration tensor is obtained by applying a unit strain on each component while fixing all of the other components as zero.

### 3.2 Modified exterior Eshelby tensor and modified exterior strain concentration tensor

The modified Eshelby tensor predicted by our analytical model is shown in Figure 4. Compared to the FEA results, the theoretical predictions are in good agreement for $S^{\mathrm{D}}_{1111}$ and $S^{\mathrm{D}}_{1212}$. The Eshelby tensor for perfect bonding is also plotted using Eqs. (4) and (5) for comparison with the modified Eshelby tensor. The absolute value of the modified exterior Eshelby tensor decreases with increasing interfacial damage, because the interaction between the inclusion and the matrix is weakened by the spring at the interface. Hence, less strain is transferred from the inclusion to the matrix, and some fraction of the energy is stored in the spring. Therefore, as $\gamma$ approaches infinity, the modified exterior Eshelby tensor approaches zero and the interior tensor converges to the identity tensor, as discussed in our previous study[29].

The modified exterior strain concentration tensor is in good agreement with the FEA

results, and the difference from the perfect bonding case can also be observed, as expected (see Figure 5). For example, for $A_{1111}^{\text{Ext,P}}$, the right and left sides near the spherical inhomogeneity have larger values than the upper and lower sides, whereas the upper and lower sides have the larger values for $A_{1111}^{\text{Ext,D}}$. The results for $A_{2211}^{\text{Ext,D}}$ and $A_{1212}^{\text{Ext,D}}$ also show the change in the distribution wherein the regions with large values in the perfect bonding case become regions with small values in the interfacial damage case.

To investigate the origin of this change, 3D isosurfaces of the modified exterior strain concentration tensor were plotted using identical values. When $\mu_0\gamma/R = 0.08$, the size of the surface is smaller than that of the perfect interface case (see Figure 6) because the bonding between the matrix and the inhomogeneity is weakened due to the finite spring compliance. The spatial distribution and magnitude are dramatically changed as gamma increases $(\mu_0\gamma/R = 0.8)$, and the isosurface becomes similar to the isosurface with a soft inhomogeneity $(E_1 = 200\text{GPa}, \nu_1 = 0.25)$ embedded in a stiff matrix $(E_0 = 400\text{GPa}, \nu_0 = 0.25)$ with perfect bonding. These results indicate that a stiffer inhomogeneity in a softer matrix does not guarantee the stiffening of composites when interfacial damage is present. In other words, when the interfacial damage is significant, the effective moduli of the composite can be inferior to that of the pure matrix.

### 3.3 Modified exterior stress concentration tensor with different interface conditions

As mentioned in Section 1~~an earlier section~~, unphysical overlapping between the matrix and inclusion (or inhomogeneity) arises in the presence of finite normal spring compliance[44, 45]. Despite this limitation, the effect of interfacial damage on the Eshelby tensor and effective moduli of composites can be described phenomenologically, and it is necessary to quantify the artifacts arising from this unphysical assumption to assess its applicability. This considers the difference in the modified stress concentration obtained with two interface conditions: one with the linear spring model, and the other with the linear spring model augmented with a non-overlapping matrix–inhomogeneity contact constraint with FEA (see Figure 7). The former is predicted using the theoretical approach and the latter is obtained with ABAQUS[46]. The modified external stress concentration tensor ($\boldsymbol{B}^{\mathrm{Ext,D}}$) was compared because observation of the overlapping phenomenon is facilitated by considering $\boldsymbol{B}^{\mathrm{Ext,D}}$ rather than $\boldsymbol{A}^{\mathrm{Ext,D}}$. For example, a uniaxial stress ($\sigma_{11}^{0}$) is applied for the prediction of $\boldsymbol{B}_{1111}^{\mathrm{Ext,D}}$. This induces negative $\varepsilon_{22}$ and $\varepsilon_{33}$ within the matrix due to the Poisson effect, and the matrix–inhomogeneity interface overlaps accordingly. In contrast, the $\varepsilon_{11}^{0}$ component of the strain is applied while fixing $\varepsilon_{22}^{0}$ and $\varepsilon_{33}^{0}$ as zero for prediction of $\boldsymbol{A}_{1111}^{\mathrm{Ext,D}}$, which does not induce negative $\varepsilon_{22}$ and $\varepsilon_{33}$ ($\sigma_{22}^{0}$ and $\sigma_{33}^{0}$ are positive because of the fixed displacement boundary condition). The modified stress concentration tensor can be obtained easily from the modified strain concentration tensor.

**Matrix**

$$\begin{aligned} \boldsymbol{L}_0^{-1}:\boldsymbol{\sigma}(\boldsymbol{x}) &= \boldsymbol{A}^{\mathrm{Ext,D}}(\boldsymbol{x}):\boldsymbol{L}_0^{-1}:\boldsymbol{\sigma}^0 \\ \boldsymbol{\sigma}(\boldsymbol{x}) &= [\boldsymbol{L}_0:\boldsymbol{A}^{\mathrm{Ext,D}}(\boldsymbol{x}):\boldsymbol{L}_0^{-1}]:\boldsymbol{\sigma}^0 \\ &= \boldsymbol{B}^{\mathrm{Ext,D}}(\boldsymbol{x}):\boldsymbol{\sigma}^0 \end{aligned}$$

(36)

**Inhomogeneity**

$$\begin{aligned} \boldsymbol{L}_1^{-1}:\overline{\boldsymbol{\sigma}^1} &= \boldsymbol{A}^{\mathrm{Int,D}}:\boldsymbol{L}_0^{-1}:\boldsymbol{\sigma}^0 \\ \overline{\boldsymbol{\sigma}^1} &= [\boldsymbol{L}_1:\boldsymbol{A}^{\mathrm{Int,D}}:\boldsymbol{L}_0^{-1}]:\boldsymbol{\sigma}^0 \\ &= \boldsymbol{B}^{\mathrm{Int,D}}:\boldsymbol{\sigma}^0 \end{aligned}$$

The difference between the two interface conditions is maximal at the limit of very large interfacial damage because the interface linear model approaches the limiting case of a porous matrix with spherical holes, while the augmented contact constraint would induce a large stress concentration in the interfacial region under compressive traction. Considering an interfacial spring compliance of $\mu_0\gamma/R = 800$, at which the modified interior strain concentration tensors, $A_{1111}^{\mathrm{Int,D}}$ and $A_{1212}^{\mathrm{Int,D}}$, become almost zero (see Appendix), the exterior stress concentration tensor is compared for two interfacial conditions. Two representative components of the modified stress concentration tensor in the $xy$-plane are visualized in Figure 8 with 2D iso-contours. The difference in the $B_{1111}^{\mathrm{Ext,D}}$ component is small because $\sigma_{11}$ near the interface is increased more by the stress concentration effect than the contact force. In contrast, the $B_{2211}$ component exhibits a noticeable difference due to the highly concentrated contact force at the upper and lower parts of the inhomogeneity (see Figure 8). Still, one can note that the simple interfacial spring model effectively captures the characteristic orientation dependence of the stress distribution around the inhomogeneity, which is important when considering the failure mode or inhomogeneity–inhomogeneity interactions of alloys and composites.

**3.4 Effective modulus of particle-reinforced composites**

The applicability of the interface spring model is also considered by calculating the effective modulus of a particle-reinforced composite based on three different approaches: the analytical prediction model in Eq. (35), FEA with a representative volume element (RVE) to validate the theoretical prediction with the interface spring condition (using COMSOL software[43]) (see Figure 7), and FEA augmented with the non-overlapping contact constraint (using ABAQUS software[46]) with the same RVE as in the second approach. All of the FEA calculations are carried out using ten independent RVE to obtain statistically meaningful results. Approximately 1,000,000 and 700,000 linear tetrahedron elements were used for the matrix and particles, respectively, and the effective modulus is obtained by averaging the normal stress at the loading surface after applying a normal strain of 0.1%.

The effective modulus of the composite is predicted for a wide range of interface spring compliance values for the combination of two particle volume fractions (5% and 10%) and two modulus ratios (2 and 10). The theoretical predictions reproduce the perfect bonding results as $\gamma$ approaches zero. When $\gamma$ increases infinitely, the modulus converges to that of the porous matrix, which can be obtained by using $\boldsymbol{L}_1 = 0$ and $\gamma = 0$ in Eq. (35) (see Figure 9). The theoretical prediction is in good agreement with the FEA result employing interface spring compliance without the additional contact constraint, as expected. In the small interfacial damage regime, the composite with a 10% particle volume fraction has a higher effective modulus than the 5% volume fraction due to the stiff embedded particles. However, in the large interfacial damage regime where this stiffening effect is decreased, the composite with the higher volume fraction has a lower modulus.

A higher effective modulus is predicted with the FEA considering the additional non-overlapping matrix–particle contact constraint. The effect of the additional contact constraint is negligible when the interfacial damage is small, whereas it becomes notable when the interfacial damage, modulus ratio, or volume fraction increases (see Figure 9). Under tension along the $x$-direction, the interaction between the matrix and particle via the contact force, noticeable in the $\sigma_{yy}$ distribution, stiffens the composite compared to its counterpart without the contact constraint, as shown in Figure 9 (e) and (f). Still, the effect of the additional contact constraint is significantly smaller than the effect of interfacial damage, and the overall behavior is captured well by the theoretical model. Specifically, the relative difference in the effective modulus at a large interfacial damage ($\mu_0\gamma/R = 800$) between the theoretical prediction and the FEA considering the contact constraint is plotted in Figure 9 (d). The relative differences are less than 2%, which implies that the theoretical approach based on the simple interface spring model is applicable for a wide range of interfacial damage.

## 4. Conclusion

Closed-form solutions for the modified exterior Eshelby tensor, strain concentration tensor, and effective modulus of the composite were derived using the interface spring model. These quantities were calculated by employing the Mandel notation to efficiently conduct fourth-order tensor operations such as inverse and double contraction, and the theoretical predictions were validated against FEA. The effect of the unphysical overlapping problem was also investigated by computing the modified stress concentration tensor and effective modulus of a particle-reinforced composite using FEA augmented with a non-overlapping contact constraint. The results show that the characteristic orientation dependence of the external stress tensor and effective modulus from the theory does not change significantly due to the additional contact condition, which indicates the applicability of the interface spring model for phenomenologically predicting the properties of composites under a wide range of interfacial damage.

**Figures and captions**

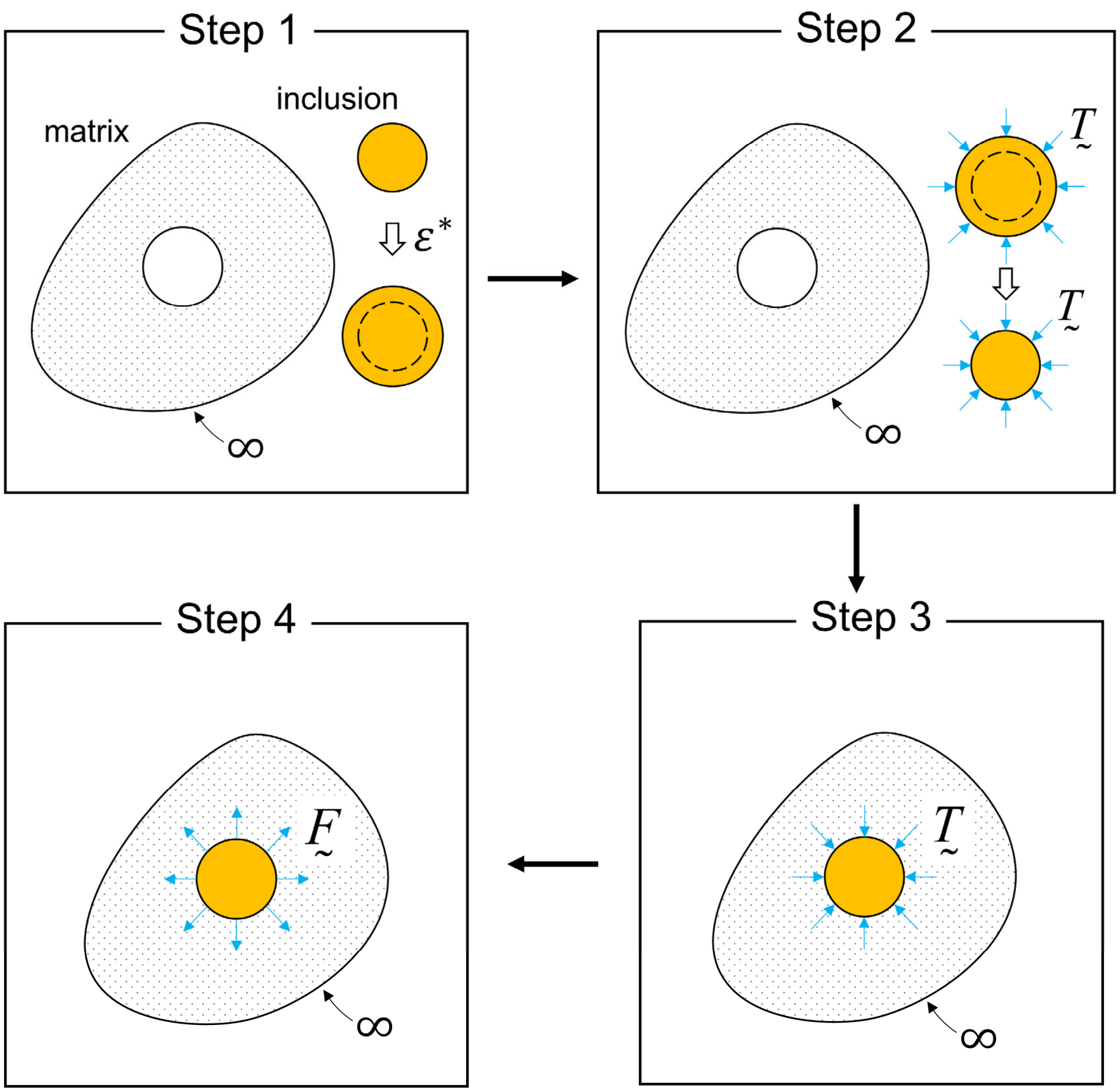

**Figure 1.** Schematic of the single inclusion problem in four steps; the inclusion is same material as the matrix, but is depicted in a different color for clarity.

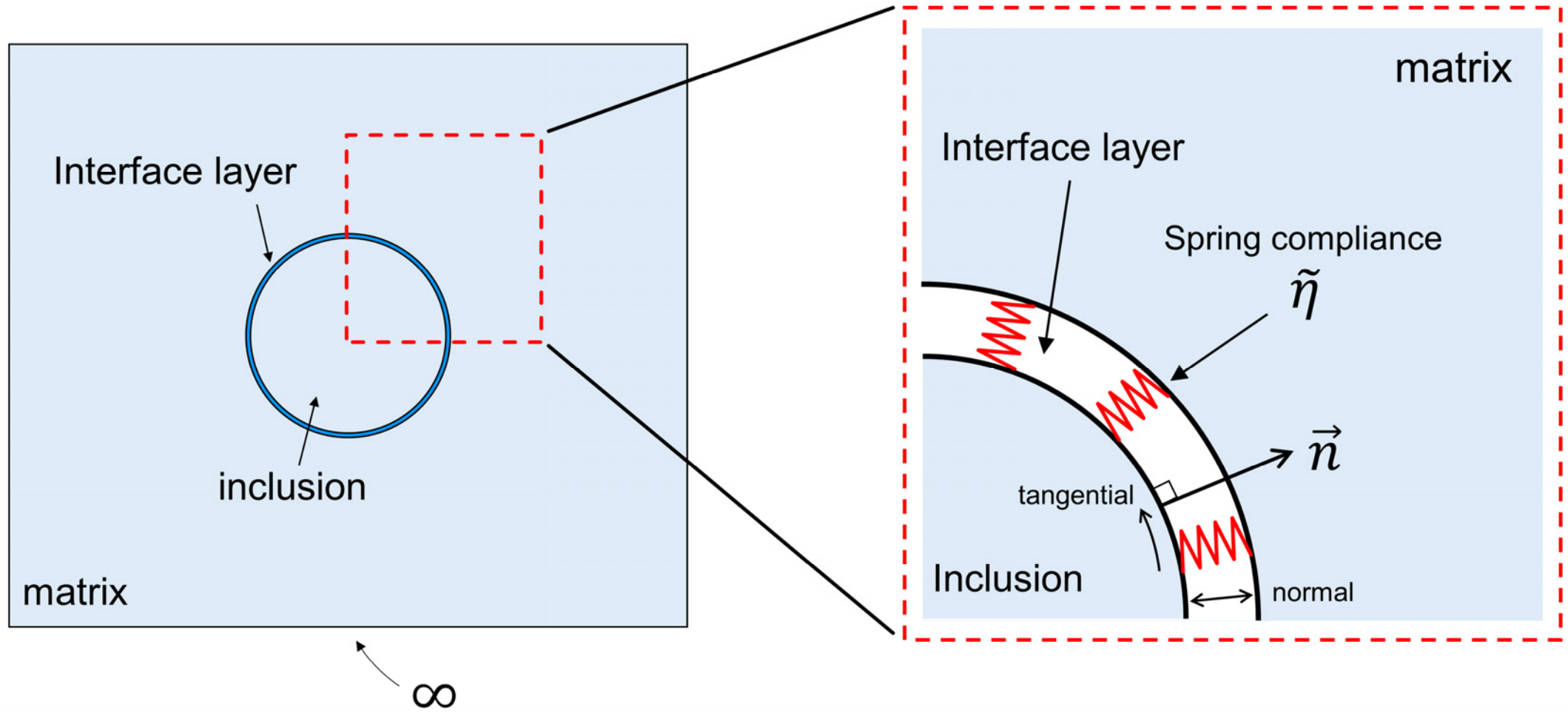

**Figure 2.** Schematic of the interface spring model; a deformed state is shown, and thus the interface has a finite thickness through a displacement jump in the tangential and normal directions.

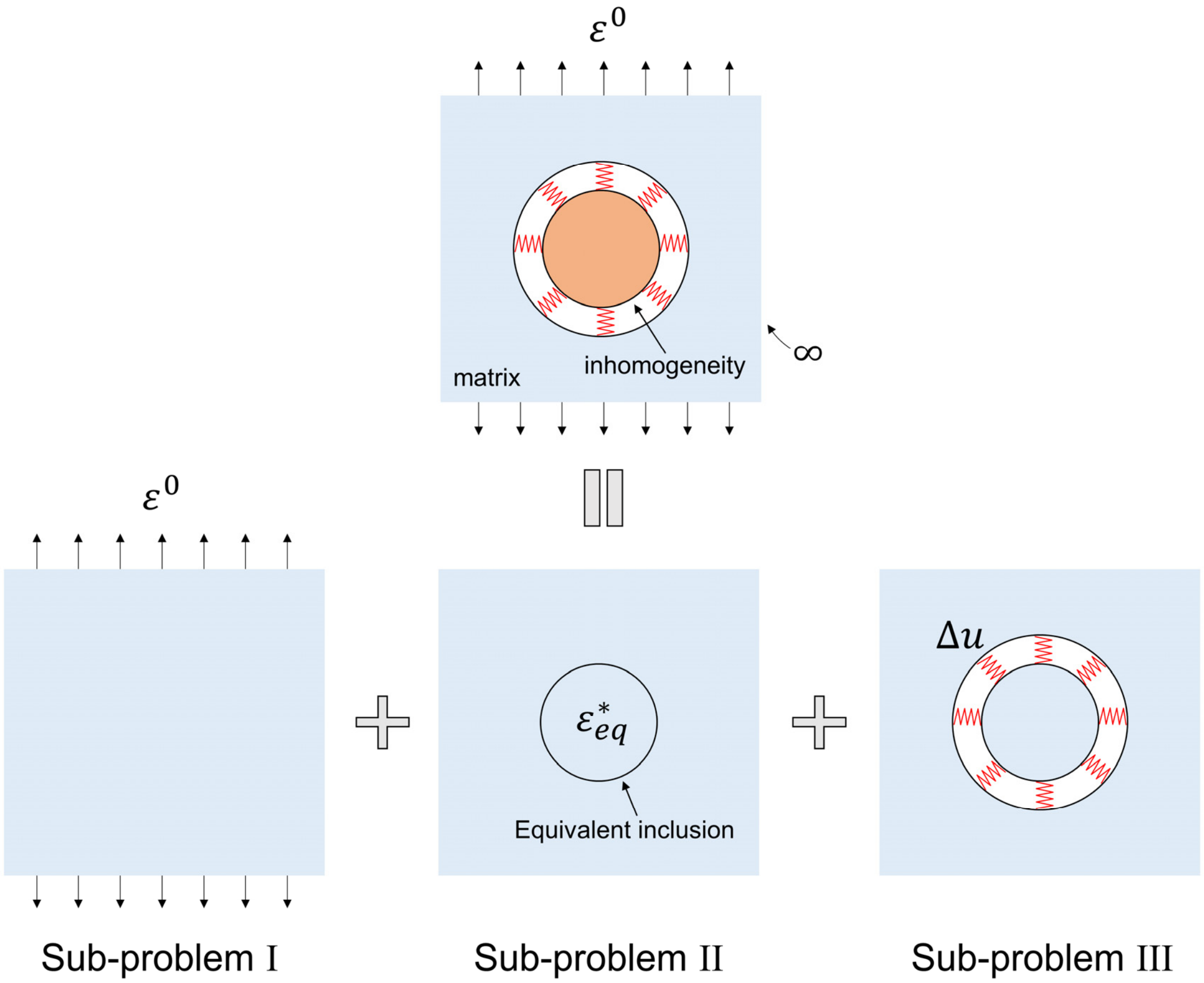

**Figure 3.** Schematic of the single inhomogeneity problem decomposed into three sub-problems

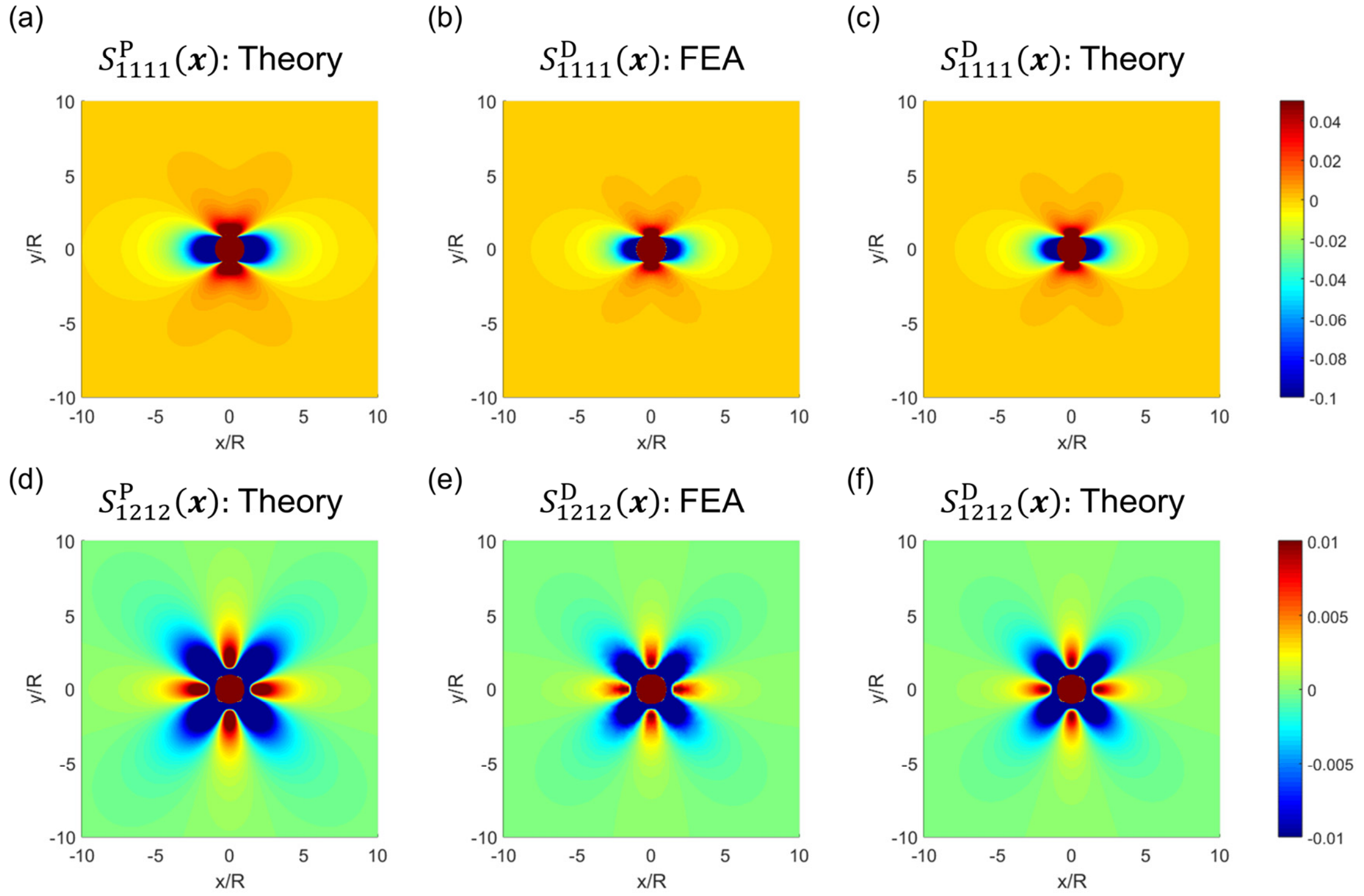

**Figure 4.** Eshelby tensor at a perfect interface ((a) $S^{\mathrm{P}}_{1111}$, (d) $S^{\mathrm{P}}_{1212}$); modified Eshelby tensor obtained with the FEA ((b) $S^{\mathrm{D}}_{1111}$ ,(e) $S^{\mathrm{D}}_{1212}$); and the theoretical prediction ((c) $S^{\mathrm{D}}_{1111}$, (f) $S^{\mathrm{D}}_{1212}$). The interfacial spring compliance used for the calculation is $\mu_0\gamma/R = 0.8$.

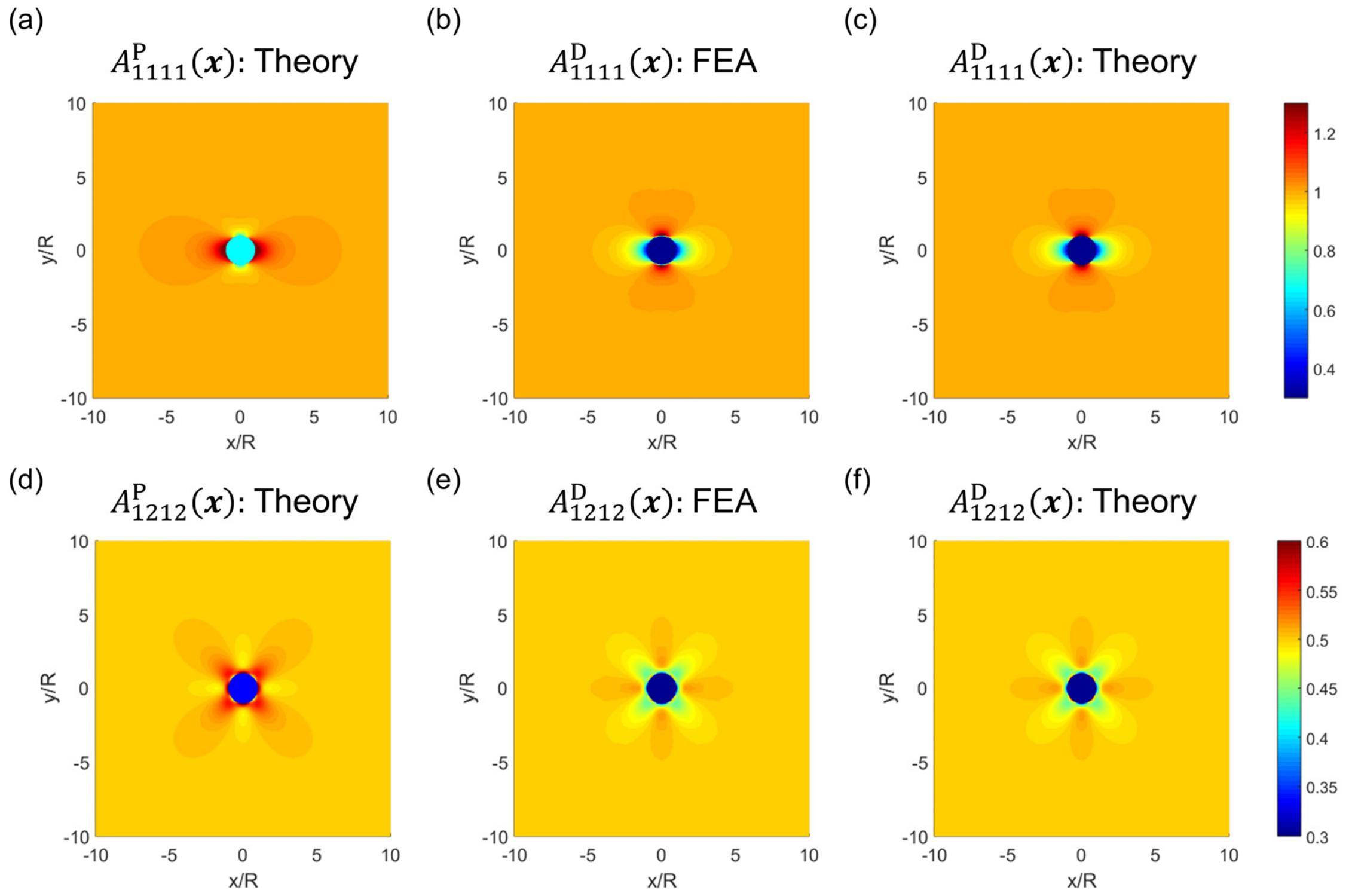

**Figure 5.** Strain concentration tensor for a perfect interface condition ((a) $A^{\mathrm{P}}_{1111}$, (d) $A^{\mathrm{P}}_{1212}$); modified strain concentration tensor with $\mu_0\gamma/R = 0.8$ obtained from the FEA ((b) $A^{\mathrm{D}}_{1111}$, (e) $A^{\mathrm{D}}_{1212}$) and theory ((c) $A^{\mathrm{D}}_{1111}$, (f) $A^{\mathrm{D}}_{1212}$)

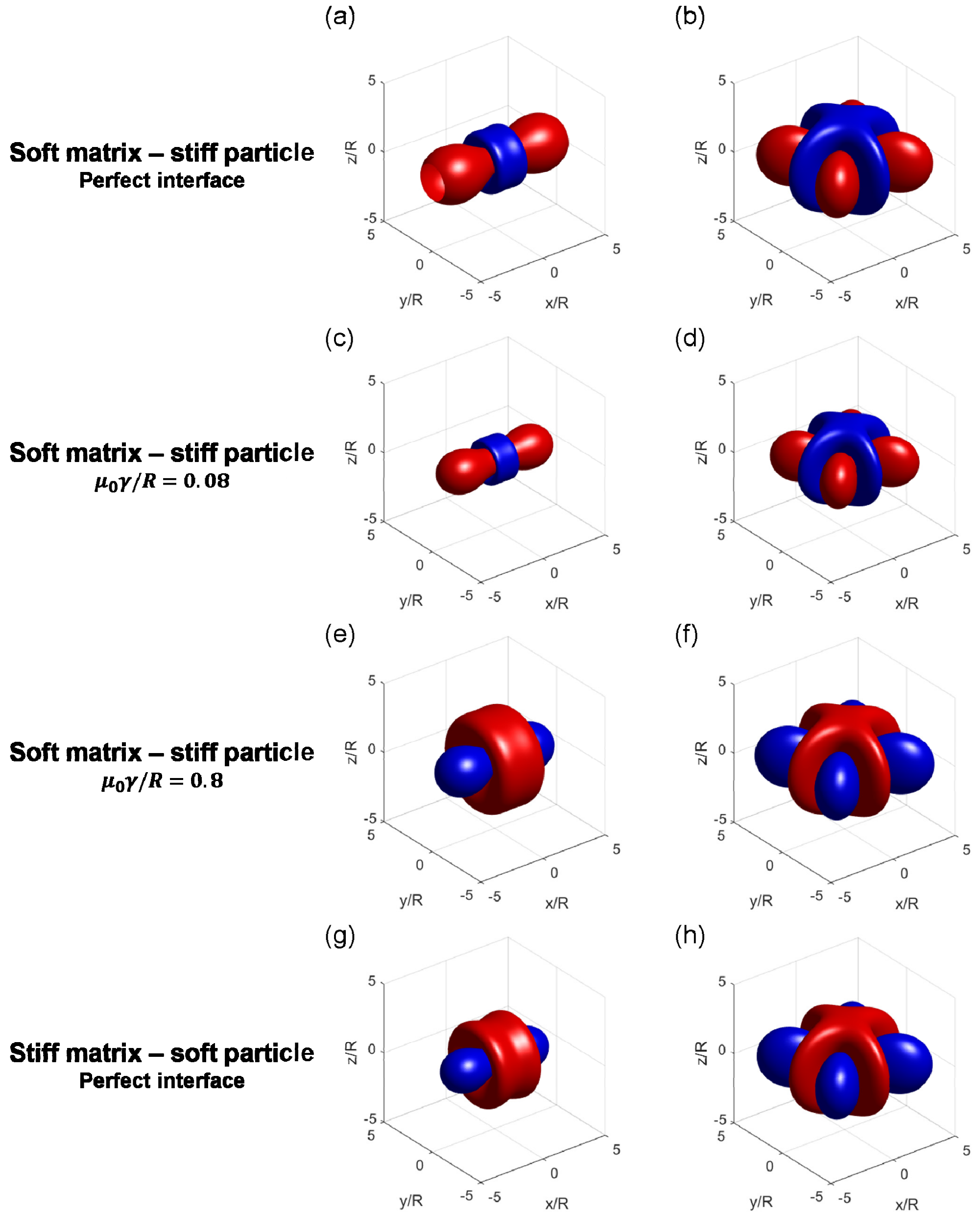

**Figure 6.** 3D isosurfaces of the modified strain concentration tensor with a perfect interface ((a) $A^{\mathrm{P}}_{1111}$ (b) $A^{\mathrm{P}}_{1212}$) and with different interfacial damage ((c),(e) $A^{\mathrm{D}}_{1111}$ (d),(f) $A^{\mathrm{D}}_{1212}$); (g) and (h) are the isosurfaces of $A^{\mathrm{P}}_{1111}$ and $A^{\mathrm{P}}_{1212}$ when the material properties of the matrix and inhomogeneity are exchanged, i.e., a stiff matrix with a soft inhomogeneity. Two constant values are plotted: 1.006 (red) and 0.985 (blue) for $A_{1111}$, and 0.502 (red) and 0.498 (blue) for $A_{1212}$.

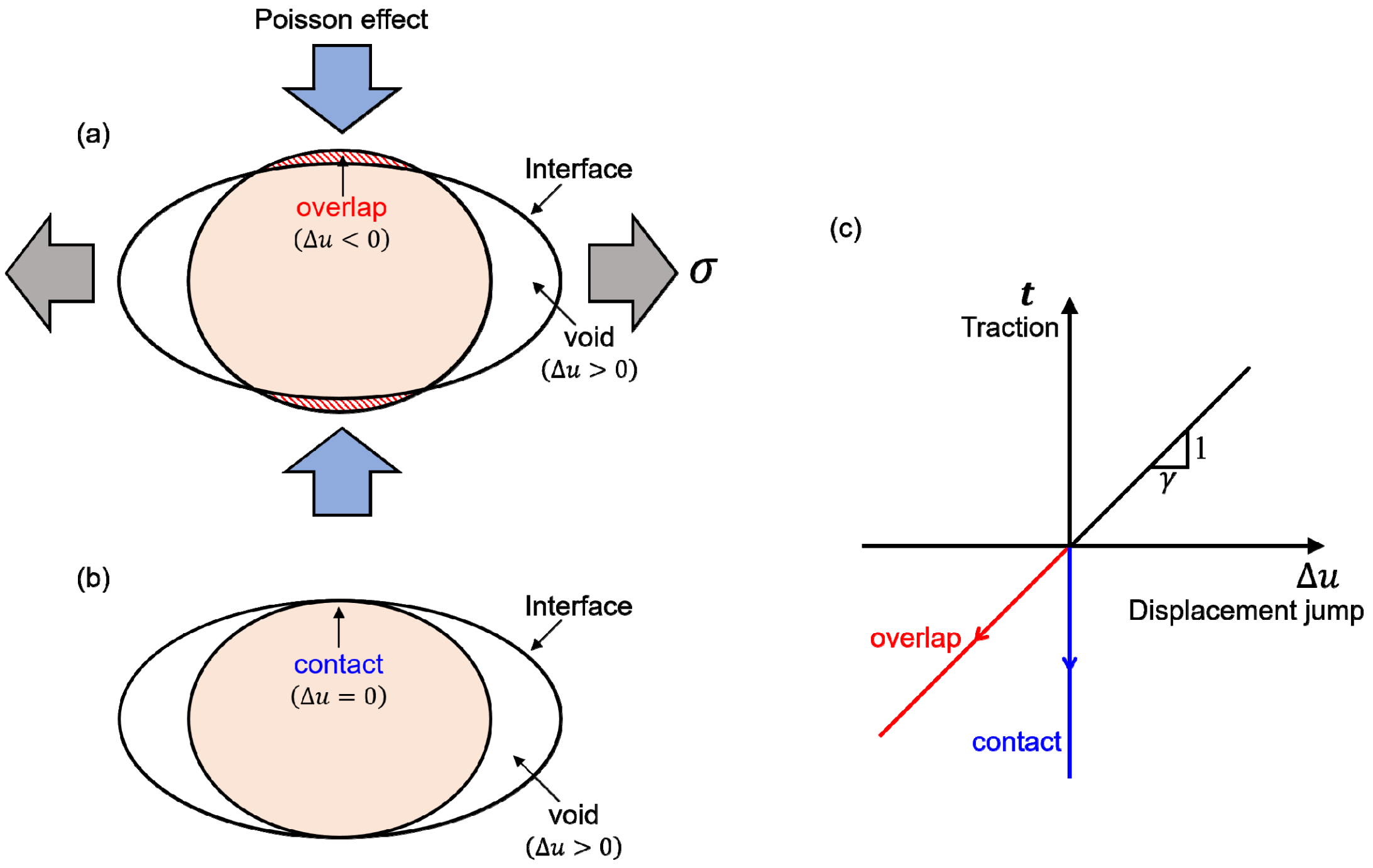

**Figure 7.** Schematic of two interface conditions: interface spring (a) allowing overlapping and (b) augmented with a contact constraint under negative traction; (c) Traction–displacement jump for the two different interface conditions (the two conditions are under the same positive traction at the interface)

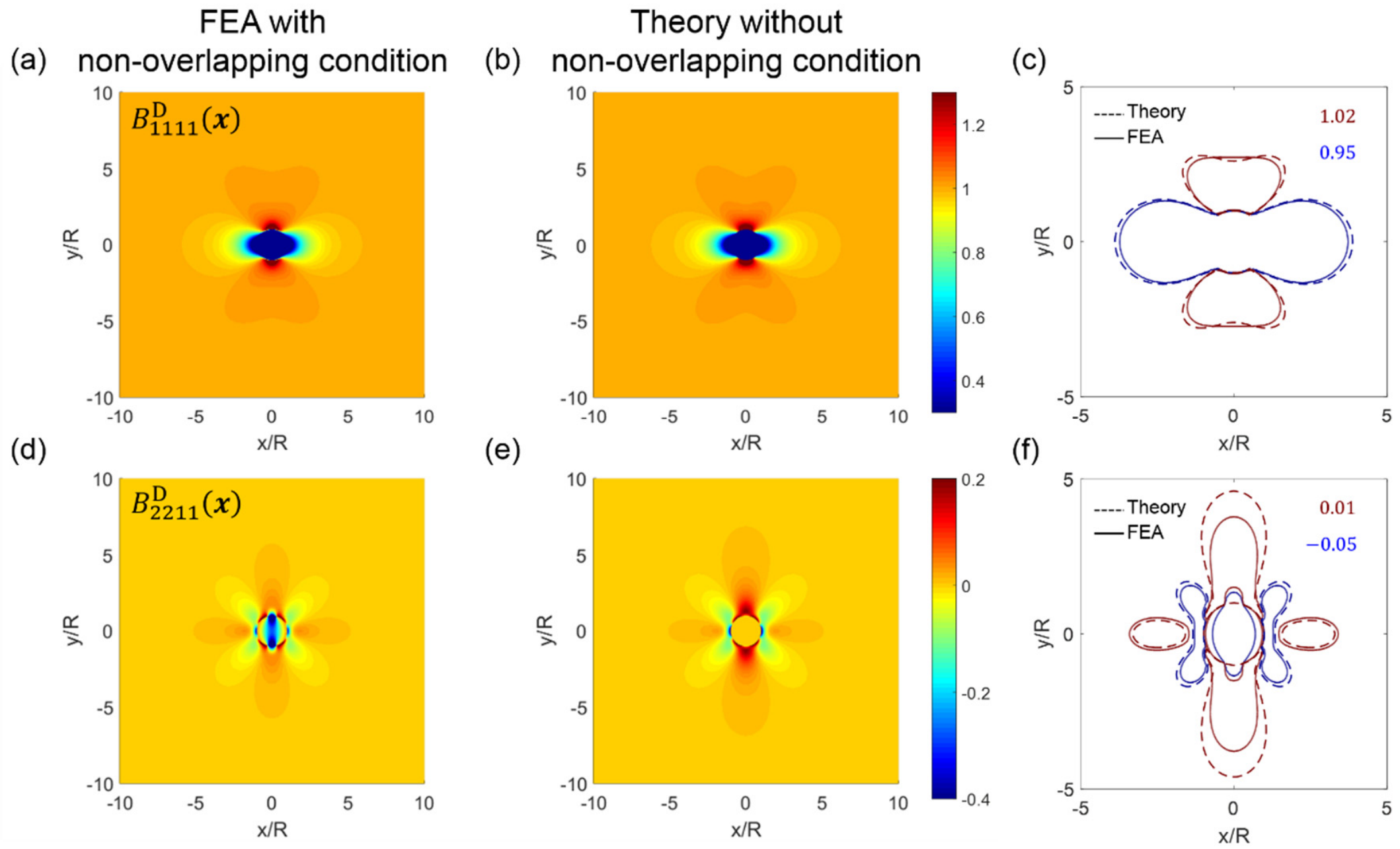

**Figure 8.** Modified stress concentration tensor calculated from the FEA ((a) $B^{\mathrm{D}}_{1111}$, (d) $B^{\mathrm{D}}_{2211}$) and theory ((b) $B^{\mathrm{D}}_{1111}$, (e) $B^{\mathrm{D}}_{2211}$). The FEA uses a non-overlapping contact boundary condition, while the theoretical approach allows overlapping at the interface. 2D iso-contours are shown for (c) $B^{\mathrm{D}}_{1111}$ and (f) $B^{\mathrm{D}}_{2211}$. The interfacial damage parameter used for these calculations is $\mu_0\gamma/R = 800$.

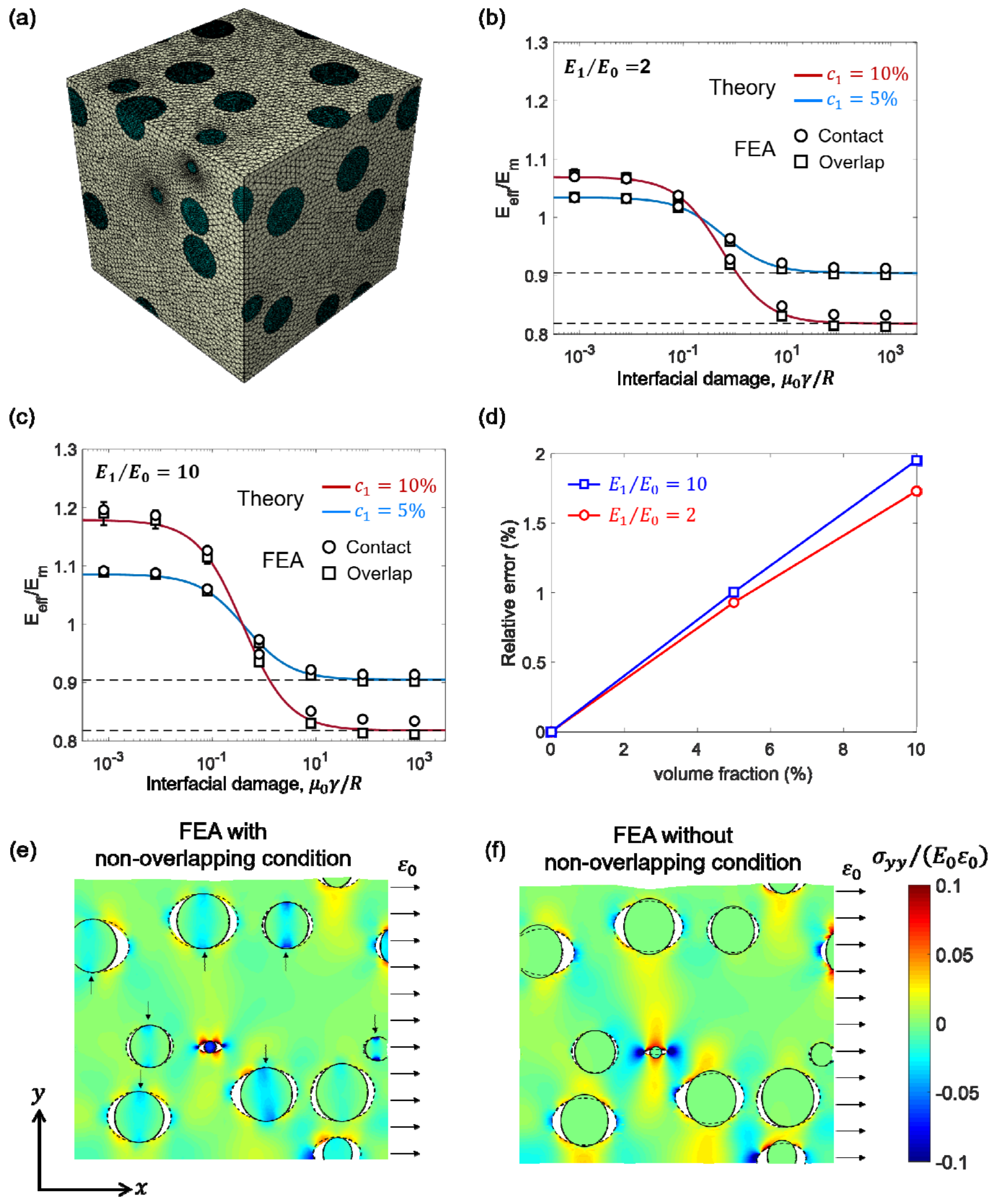

**Figure 9.** (a) Particle-reinforced composite RVE used for the FEA. Effective Young's modulus of the composite for matrix–particle modulus ratios $(E_1/E_0, E_0 = 200\ \text{GPa}, \nu_0 = \nu_1 = 0.25)$ of (b) 2 and (c) 10 over a wide range of interfacial damage. Because a sufficiently large RVE was used, some of the error bars are even smaller than the size of the mark. (d) Relative error obtained by comparing the theoretical prediction and FEA result considering contact at $\mu_0\gamma/R = 800$. Stress contours under $\varepsilon_{xx}$ loading and two interface conditions ((e) contact and (f) overlapping) at $\mu_0\gamma/R = 800$. The volume fraction is 10% with $E_1/E_0 = 2$, and the same RVE is used for the calculations.

**Appendix**. **Modified interior strain concentration tensor**

The modified interior strain concentration was calculated for a wide range of interfacial damage (see Figure 10). It converges to zero as $\gamma$ increases because the strain from the far field strain is decreased due to the weakened interface. Therefore, as $\gamma$ increases, the strain at the matrix approaches the elasticity solution for an infinite medium with a spherical hole, and it can be predicted using Eq. (33) with $\gamma = 0$ and $\boldsymbol{L}_1 = 0$.

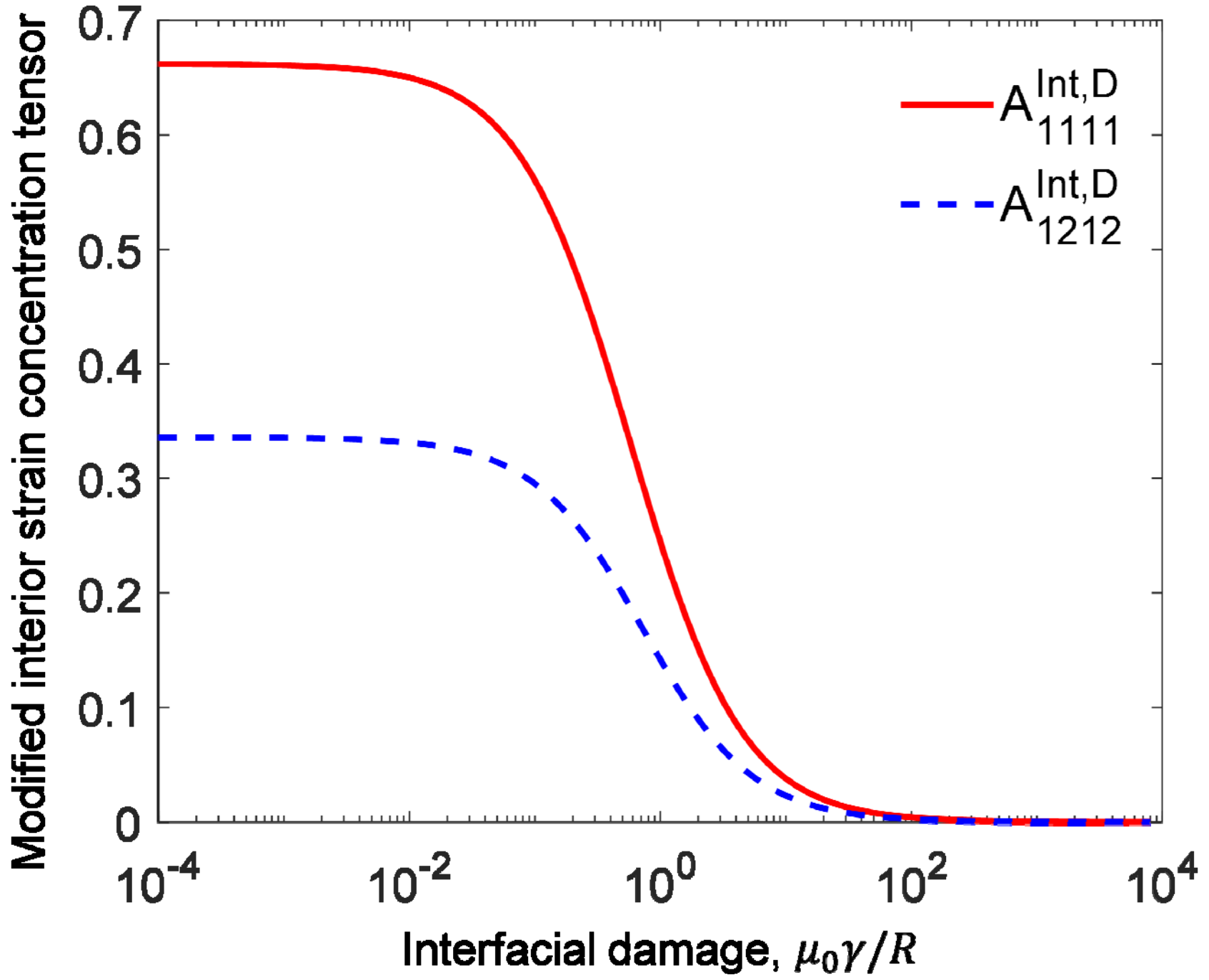

**Figure 10.** Modified interior strain concentration tensor with respect to interface spring compliance